\documentclass[preprint]{revtex4}

\usepackage{amssymb}
\usepackage{amsmath}
\usepackage{graphicx}
\usepackage{multirow}
\usepackage{caption}
\usepackage{subfig}
\begin{document}
\title
{Stochastically driven instability in rotating shear flows}
\author
{Banibrata Mukhopadhyay$^{1}$ and Amit K. Chattopadhyay$^{2}$
\\
1. 
Department of Physics, Indian Institute of Science, Bangalore 560 012, India;\\
bm@physics.iisc.ernet.in \\
2. Aston University, Non-linearity and Complexity Research Group, Engineering \\
and Applied Science, Birmingham B4 7ET, UK; a.k.chattopadhyay@aston.ac.uk
}


\begin{abstract}

Origin of hydrodynamic turbulence in rotating shear flows is investigated. The particular emphasis
is the flows whose angular velocity decreases but specific angular momentum increases with increasing
radial coordinate. Such flows are Rayleigh stable, but must be turbulent in 
order to explain observed data. 
Such a mismatch between the linear theory and observations/experiments is more severe
when any hydromagnetic/magnetohydrodynamic instability 
and then the corresponding turbulence therein is ruled out. 
The present work explores the effect of stochastic noise on such hydrodynamic flows. We essentially
concentrate on a small section of such a flow which is nothing but a plane shear flow supplemented by 
the Coriolis effect. This also mimics a small section of an astrophysical accretion disk.
It is found that such stochastically driven flows exhibit large temporal and 
spatial correlations of perturbation velocities, and hence large energy 
dissipations of perturbation, which
presumably generate instability. 
A range of angular velocity ($\Omega$) profiles of background flow, starting from that of constant 
specific angular momentum ($\lambda=\Omega r^2$; $r$ being the radial coordinate) to that
of constant circular velocity ($v_\phi=\Omega r$), is explored. 
However, all the background angular velocities exhibit identical growth and roughness exponents of perturbations, revealing
a unique universality class for the stochastically forced hydrodynamics of rotating shear flows.
This work, to the best of our knowledge, is the first attempt to understand origin of instability
and turbulence in the three-dimensional Rayleigh stable rotating shear flows by introducing
additive noise to the underlying linearized governing equations.
This has important implications to resolve the turbulence 
problem in astrophysical hydrodynamic flows such as accretion disks.

\end{abstract}

\maketitle

\textit{Keywords}: 
hydrodynamics; instabilities; turbulence; statistical mechanics; accretion, accretion disks

\vspace{0.2cm}

\textit{PACS}: 47.85.Dh, 95.30.Lz; 47.20.Ib; 47.27.Cn; 05.20.Jj; 98.62.Mw

\section{Introduction}


There are certain rotating shear flows which are Rayleigh stable
but experimental/observational data argue them to be turbulent.  
In absence of magnetic coupling, such flows are stable under linear perturbation. 
What drives their instability and then turbulence
in absence of linearly unstable modes?
In order to understand a
plausible route to hydrodynamic turbulence in such flows, a series of papers 
was published by different independent groups
\cite{chage,tev,yeko,man,amn}. Based on
``bypass mechanism" \cite{butler,reddy,trefethen}, those papers showed that such flows
exhibit large transient energy growth
of linear perturbation. It was argued earlier that indeed the large transient energy growth 
is responsible for subcritical transition to turbulence in plane Couette flow and 
plane Poiseuille flow, as seen in laboratory \cite{farrell}. In such flows the transient growth increases
with increasing Reynolds number ($R_e$). For example, in plane Couette flow, the maximum 
transient growth scales as $R_e^2$. However, the situation changes drastically when
rotational effects come into the picture. The Coriolis force is the main culprit 
behind this change. It was shown earlier \cite{man,amn,bmraha} that rate of the increase of transient growth
in two-dimensional (insignificant vertical scale height) linearized rotating shear flows 
with Keplerian angular velocity ($\Omega$)
profile ($\Omega\sim r^{-3/2}$; $r$ being the radial coordinate of the flow) decreases
much compared to that in plane Couette flows, when maximum growth scales as $R_e^{2/3}$. Furthermore, in 
three dimensions (with finite height), such transient growth in Keplerian flows is insignificant, independent of viscosity,
to generate any instability and turbulence.
But the three-dimensional Keplerian flow is an important natural flow,
which exists in several 
astrophysical contexts. Interestingly, while some authors \cite{ji}, based on
the prototype experiment consisting of fluid confined between two corotating 
cylinders, reported the Rayleigh stable rotating shear flows, similar
to that of a Keplerian disk, laminar, some other \cite{lathrop} 
found it turbulent. However, the results from direct numerical simulations 
\cite{avila} argue that laboratory flows are indeed hydrodynamically unstable 
and should become turbulent at low Reynolds numbers. 
But the last author \cite{avila} and the present authors \cite{kanak} also showed/argued that
the experiments are compromised by undesired effects due to the finite
length of the cylinders.
Note that some (effective) nonlinear theory \cite{bm06,rajesh} revealing 
large energy growth of perturbation (even of finite amplitude) argues for
plausible routes to turbulence in three dimensions. Hence, the real challenge is to 
uncover the mystery of mismatch between theory and observation under linear theory,
in absence of a finite amplitude of perturbation.

As argued earlier \cite{bmraha} that in three dimensions, one requires to invoke extra physics,
e.g. external or self-generated noise etc., to 
reveal large energy growth or even instability in the system.
It is needless to mention that the real flows
must be three-dimensional, e.g. an astrophysical accretion disk must have a thickness, however small may be,
and hence rotation plays a significant role to determine the dynamics of the flow.
The aim of the present work is to investigate the amplification of linear perturbations
which may eventually lead to hydrodynamic instability and then plausible turbulence 
in certain rotating shear flows supplemented by stochastic fluctuations in three dimensions.

There is a long term association of growing, unstable modes generated by perturbed 
flows with statistical physics. The flow fields perturbed either at the boundary or through external forcing
have been shown to produce emergent instabilities \cite{chandra81,batchelor00}, altering the scaling 
structure of the systems. 
It was shown by Forster, Nelson and Stephen \cite{nelson} that in the long time, large distance 
asymptotic limit, large non-equilibrium fluctuations in unbounded flows decouple to stabilize the flow. 
The authors also studied the critical dimension beyond which fluctuations get redundant. However 
below that dimension,
all the Brownian oscillations perturb the flow in the equilibrium limit. In a follow-up work by 
De Dominicis and Martin \cite{dedominicis_martin}, 
a further generalization of such `forced Navier-Stokes flows' was provided incorporating a range of 
possibilities for the perturbing random forces which is correlated with the Kolmogorov spectrum 
\cite{kolmogorov1941,kolmogorov1962} to redefine the scaling laws in the presence of such stochastic 
components. Later on, the focus was on studying topical variations in the transport properties of fluids under 
stochastic perturbations \cite{jkb96}. These works were the precursor to the Kardar-Parisi-Zhang 
(KPZ) model \cite{kpz} which studied a differential variant of the Burger field to illumine fundamental 
hydrodynamic instabilities related to the intermittency spectrum \cite{chekhlov_yakhot,akc_epl}.
The KPZ model also sheds 
light on varieties of non-equilibrium phenomena, including fluid flows, bacterial growth, paper burning 
front, etc. (see, \cite{barabasi_stanley}). All such results were later encapsulated 
in another illustrative work by Medina, Kardar, Hwa and Zhang \cite{medina_kardar_hwa_zhang} 
where the effects of varying spatio-temporal correlations were studied on the Burger equation 
(which is a one-dimensional variant of the Navier-Stokes equation). It also showed for the first
time that the statistical 
field theory could be a legitimate consort of fluid mechanics such that local instabilities seen in fluid flows 
could often be identified as non-trivial infra-red divergences. All the studies, however,
were restricted mostly to trivial/periodic and open/absorbing boundary conditions.   

Chattopadhyay \& Bhattacharjee \cite{akc_shear} applied the randomly stirred model of Navier-Stokes flows 
constrained by geometry
in a flow involving a three-dimensional {\it bounded} incompressible 
fluid perturbed by boundary layer shear. 
In that work, they considered two relatively shearing flat plates with a layer of fluid sandwiched in between. 
The effect of the constrained flow is implemented through the stochasticity structure function, which 
ensures that the symmetry is violated through noise autocorrelation function into the problem. 
Note that the geometry could also be
changed over to rotating co-axial cylinders \cite{landau87} for externally impressed shearing forces,
essentially bearing the same qualitative imprint. 
This problem, known as boundary layer turbulence, has recently been revisited experimentally in 
laboratory, by generating shear in the flow through a raft 
of blowing air \cite{KITP}. A numerical simulation governing large Eddy also reveals
same experimentally observed non-inertial effects \cite{meneveau}. 
Note in a related context that astrophysical observed data indeed argued for the signature of
chaos in rotating shear flows, more precisely accretion flows \cite{chaos1,chaos2}, which further 
supports infall of matter towards black holes \cite{rm10} and neutron stars \cite{mg03} 
due to turbulent viscosity.

In the present study, we implement the ideas of statistical physics discussed above
to rotating shear flows in order to obtain the correlation
energy growth of fluctuation/perturbation and underlying scaling properties. 
In the next section, we discuss the equations describing the stochastically forced perturbed
flows which are to be solved for the present purpose. 
Subsequently, in \S 3 we investigate
the temporal and spatial correlations of perturbation in detail, in order to understand the
plausible instability in the flows. Finally, we summarize the results with conclusions
in \S 4.

\section{Equations describing forced perturbed accretion flow}

As our specific interest lies in hydrodynamic instability and turbulence 
in presence of stochastic noise, we straight away recall the Orr-Sommerfeld and Squire equations 
in presence of the Coriolis force \cite{man,amn,bmraha}, but supplemented by
stochastic noise.
They are established by eliminating pressure from the linearized Navier-Stokes equation with
background plane shear $(0,-x,0)$ 
neglecting magnetic coupling in presence of angular velocity 
$\Omega\sim r^{-q}$ in a small section of the incompressible flow \footnote{A set of
magnetized version of the equations is given in the Appendix.}. Hence, the underlying equations
are nothing but the linearized set of hydrodynamic equations combining with the equation of continuity in 
local Cartesian coordinates ($x,y,z$) at an arbitrary time $t$ (see, e.g., \cite{man,bmraha} for detailed description of the
choice of coordinate in a small section) given by
%
\begin{equation}
\left(\frac{\partial}{\partial t}-x\frac{\partial}{\partial y}\right)\nabla^2 u
+\frac{2}{q}\frac{\partial \zeta}{\partial z}
=\frac{1}{R_e}\nabla^4 u+\eta_1(x,t),
\label{orrv}
\end{equation}
\begin{equation}
\left(\frac{\partial}{\partial t}-x\frac{\partial}{\partial y}\right)\zeta
+\frac{\partial u}{\partial z}
-\frac{2}{q}\frac{\partial u}{\partial z}=\frac{1}{R_e}\nabla^2 \zeta +
\eta_2(x,t),
\label{zeta}
\end{equation}
where $u,\zeta$ represent respectively $x-$components of velocity and vorticity perturbation, 
$R_e$ is the Reynolds number, $\eta_{1,2}$ are the components of noise arising in the linearized system due
to stochastic perturbation such that $<\eta_i(\vec x,t) \eta_j(\vec x',t')>=D_i(\vec x)\:\delta^3(\vec x-\vec x')\:\delta(t-t')\:\delta_{ij}$. 
Note that equations (\ref{orrv}) and (\ref{zeta}) describe perturbations in a small section of 
an accretion flow which could be expressed in local Cartesian coordinates. 
The long time, large distance behavior of the velocity correlations are encapsulated in 
$D_i(\vec x)$ which is a structure pioneered by Forster, Nelson and Stephen \cite{nelson}. 
In the Fourier transformed space, the specific structure of the correlation function $D_i(\vec k)$
depends on the regime under consideration. It can be shown for all (non-linear) non-inertial flows 
\cite{nelson,akc_shear} that $D_i(k) \sim 1/k^d$, where $d$ is the spatial dimension, without
vertex correction and $D_i(k) \sim 1/k^{d-\alpha}$ with $\alpha>0$, in presence of vertex correction.
Note, however, that $D_i(\vec x)$ is constant for white noise.

We now focus onto the narrow gap limit which otherwise leads to an approximation 
$x\rightarrow L$, when $L$ being the (narrow) size of the shearing box in the $x$-direction. 
The appropriate choice of a sharing box has already been described 
by the previous authors; see, e.g. \cite{amn,bmraha}, which we do not repeat here again. 
In brief, in a local analysis one considers a small radially confined region of the accretion flow,
while the azimuthal confinement is imposed by a periodic boundary condition. This effectively
gives rise to a shearing Cartesian box for geometrically thin flows. 
We further choose a very small $L$ so that we are essentially interested in an orbit. 
This is in the same spirit as of the
celebrated work by Balbus \& Hawley \cite{bh91} who initiated the
existence of Magneto-Rotational-Instability (MRI) with axisymmetric perturbation. 
One of the motivations 
behind it is as follows. Presence of turbulence in shearing astrophysical accretion disks is mandatory to
generate viscosity therein. It is only the viscosity which
helps in transporting the energy-momentum from an outer orbit to an inner orbit leading to the inspiral of matter, 
giving rise to a differential rotating (shearing) structure of accretion disks. As the aim is to address the
origin of such viscosity (which can not be of molecular) for rotating shear flows, choosing $x$ to be fixed justified.

Hence, in the above discussed framework, we can resort to a Fourier series expansion of
$u$, $\zeta$ and $\eta_i$ as
\begin{eqnarray}
\nonumber
u(\vec{x},t)=\int\tilde{u}_{\vec{k},\omega}\,e^{i(\vec{k}.\vec{x}-\omega t)}d^3k\,d\omega,\\
\nonumber
\zeta(\vec{x},t)=\int\tilde{\zeta}_{\vec{k},\omega}\,e^{i(\vec{k}.\vec{x}-\omega t)}d^3k\,d\omega, \\
\eta_i(\vec x,t) = \int\tilde{\eta_i}_{\vec{k},\omega}\,e^{i(\vec{k}.\vec{x}-\omega t)}d^3k\,d\omega,
\label{four}
\end{eqnarray}
where $\vec{k}$ and $\omega$ respectively represent the wave-vector and angular frequency of perturbation,
and substituting them into equations (\ref{orrv}) and (\ref{zeta}) we obtain 
\begin{eqnarray}
\left(\begin{array}{cr}\tilde{u}_{\vec{k},\omega}\\ 
\tilde{\zeta}_{\vec{k},\omega}\end{array}\right)={\cal M}\left(\begin{array}{cr}\tilde{\eta_1}_{\vec{k},\omega}\\ 
\tilde{\eta_2}_{\vec{k},\omega}\end{array}\right),
\label{mat1}
\end{eqnarray}
where 
\begin{eqnarray}
{\cal M}=\left(\begin{array}{cr}{\cal M}_1\,\,\,\,\, {\cal M}_2\\ 
{\cal M}_3\,\,\,\,\, {\cal M}_4\end{array}\right),
\label{mat2}
\end{eqnarray}
\begin{eqnarray}
\nonumber
{\cal M}_1=\frac{q^2R_e(k^2-iR_e\omega-iLR_ek_y)}{-k^2q^2(k^2-iR_e\omega-iLR_ek_y)^2+2(q-2)R_e^2k_z^2},\\
\nonumber
{\cal M}_2=-\frac{2iqR_e^2k_z}{-k^2q^2(k^2-iR_e\omega-iLR_ek_y)^2+2(q-2)R_e^2k_z^2},\\
\nonumber
{\cal M}_3=-\frac{i(q-2)qR_e^2k_z}{-k^2q^2(k^2-iR_e\omega-iLR_ek_y)^2+2(q-2)R_e^2k_z^2},\\
{\cal M}_4=\frac{k^2q^2R_e(k^2-iR_e\omega-iLR_ek_y)}{k^2q^2(k^2-iR_e\omega-iLR_ek_y)^2-2(q-2)R_e^2k_z^2},
\label{am}
\end{eqnarray}
when $\tilde{{\eta}_{i}}_{\vec{k},\omega}$; $i=1,2$, are the components of noise in $k-\omega$ space,
$k=\sqrt{k_x^2+k_y^2+k_z^2}$.

\section{Two-point correlations of perturbation}

We now look at the spatio-temporal autocorrelations of the flow fields $u$ and $\zeta$ \cite{barabasi_stanley}. 
For the present purpose, the magnitudes and gradients (scalings) of these autocorrelation functions of perturbations would 
plausibly indicate noise induced instability which could lead to 
turbulence in rotating shear flows.

\subsection{Temporal correlation}

Assuming $<\tilde{\eta_1}_{\vec{k},\omega}\,\tilde{\eta_2}_{\vec{k},\omega}>=0$, 
we obtain the temporal correlations of velocity,
vorticity and gradient of velocity given below as
\begin{eqnarray}
\nonumber
&&<u(\vec{x},t)\,u(\vec{x},t+\tau)>=C_u(\tau)=\int d^3k\,d\omega\,e^{-i\omega\tau}<\tilde{u}_{\vec{k},\omega}\,
\tilde{u}_{-\vec{k},-\omega}>,\\
\nonumber
&&<\zeta(\vec{x},t)\,\zeta(\vec{x},t+\tau)>=C_\zeta(\tau)=\int d^3k\,d\omega\,e^{-i\omega\tau}<\tilde{\zeta}_{\vec{k},\omega}\,
\tilde{\zeta}_{-\vec{k},-\omega}>,\\
\nonumber
&&{\hskip-1.3cm <\frac{\partial u(\vec{x},t)}{\partial x}\,\frac{\partial u(\vec{x},t+\tau)}{\partial x}>
=C_{du}(\tau)
=\int d^3k\,d\omega\,e^{-i\omega\tau}k_x^2<\tilde{u}_{\vec{k},\omega}\,
\tilde{u}_{-\vec{k},-\omega}>},\\
\label{velzetdercor}
\end{eqnarray}
where $\tau$ be the time at which the correlations to be measured.
Now using equations (\ref{mat1}), (\ref{mat2}), (\ref{am}) and (\ref{velzetdercor})
\begin{eqnarray}
\nonumber
&&C_u(\tau)
=\int d^3k\,d\omega\,e^{-i\omega\tau}\,[2D_1q^4R_e^2(k^4+(R_e\omega+LR_ek_y)^2)\\
\nonumber
&&+8D_2q^2R_e^4k_z^2]/[-4(-2+q)^2R_e^4k_z^4\\
\nonumber
&&+4R_e^2q^2(-2+q)k^2k_z^2(k^4-(R_e\omega+LR_ek_y)^2)\\
&&-k^4q^4(k^4+(R_e\omega+LR_ek_y)^2)^2]. 
\label{velcor}
\end{eqnarray}

\noindent
We further consider the projected hyper-surface for which $k_x=k_y=k_z=k/\sqrt{3}$, without much loss of
generality for the present purpose. As our one of the major interests is to determine
the scaling laws, this restriction would not matter. However, this is equivalent to a 
special choice of initial perturbation which will affect the magnitude of the correlations. 
We now perform the $\omega$-integration by computing 
the poles of the kernel given by
\begin{eqnarray}
\omega_1 &=& \frac{1}{R_e}(\sqrt{\frac{2R_e^2 (2-q)}{3q^2}-k^4+(\frac{2\sqrt{2}R_e(2-q)}{\sqrt{3}q})k^2}
-\frac{LR_ek}{\sqrt{3}}), \nonumber \\
\omega_2 &=& \frac{1}{R_e}(\sqrt{\frac{2R_e^2 (2-q)}{3q^2}-k^4-(\frac{2\sqrt{2}R_e(2-q)}{\sqrt{3}q})k^2}-
\frac{LR_ek}{\sqrt{3}}), \nonumber \\
\omega_3 &=& -\frac{1}{R_e}(\sqrt{\frac{2R_e^2 (2-q)}{3q^2}-k^4+(\frac{2\sqrt{2}R_e(2-q)}{\sqrt{3}q})k^2}-
\frac{LR_ek}{\sqrt{3}}), \nonumber \\
\omega_4 &=& -\frac{1}{R_e}(\sqrt{\frac{2R_e^2 (2-q)}{3q^2}-k^4-(\frac{2\sqrt{2}R_e(2-q)}{\sqrt{3}q})k^2}-
\frac{LR_ek}{\sqrt{3}}).\nonumber \\
\label{poles}
\end{eqnarray}

By summing up the residues at the appropriate poles in presence of white noise, we evaluate the magnitude of frequency part of the 
integration given by
\begin{eqnarray}
\nonumber
&&{\rm RES}_{C_u}(q,R_e,\tau) = \frac{2\sqrt{2}\tau}{\sqrt{3(2-q)}}\frac{\pi R_e}{q k^6}\frac{1}{a^2+b^2}\times \\
\nonumber
&&[(bm-an)\sin(a)\cosh(b)+(am+bn)\cos(a)\sinh(b)], \\
&&{\rm where} \\
\nonumber
&&a = \sqrt{2\tilde{x}+\sqrt{4\tilde{x}^2+\tilde{y}^2}},\,\
b = \sqrt{\tilde{x}+\sqrt{4\tilde{x}^2+\tilde{y}^2}},\\ 
\nonumber
&&\tilde{x} = -[\frac{k^4}{R_e^2}+\frac{2}{3}(\frac{(2-q)}{q^2})]\tau^2,\,\  
\tilde{y} = \frac{2}{3}\sqrt{6(2-q)}(\frac{k^2}{q R_e})\tau^2, \\
\nonumber
&&m = 2k^2(2-q),\,\ 
n = \sqrt{6(2-q)}(\frac{qk^2}{R_e}).
\end{eqnarray}
Hence
\begin{eqnarray}
&&C_u(\tau) = 4\pi \int_{k_0}^{k_m}\:dk\: k^2 \: {\rm RES}_{C_u}(q,R_e,\tau),\\
\nonumber
&&{\rm with}\,\,\,D_1=D_2=1,
\end{eqnarray}
where $k_0=2\pi/L_{\rm max}$ and 
$k_m=2\pi/L_{\rm min}$, 
when $L=L_{\rm max}-L_{\rm min}$ 
being the size of the chosen small section of the flow in the radial direction, 
chosen to be $2$ throughout for the present calculations.
Similarly, $C_\zeta(\tau)$ and $C_{du}(\tau)$ can be easily obtained from equation 
(\ref{velzetdercor}).
Note that the choice of constant $D_1$ and $D_2$ is due to our primary 
consideration of 
linear perturbation of the flow variables in presence of an additive noise.  
However, $D_1$ and $D_2$ need not be constant (viz. \cite{nelson}) always. Later,
we will discuss the cases with $D\sim k^{-d}$ what effectively does not alter
our results.

\subsubsection{Asymptotic scaling laws}
\begin{figure}
   \centering
\includegraphics[scale=0.8]{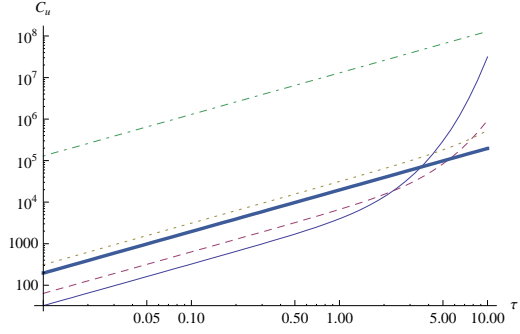}
\caption{Temporal correlation as a function of time for $R_e=10000$, when $q=1$ (solid line), $1.5$ (dashed line), $1.9$ (dotted line), $1.9999$ (dotdashed line) and
flow is non-rotating (thick line).
}
\label{tempr1e4}
\end{figure}

At a large $R_e$ ($\to \infty$) $C_u$ reduces to
\begin{subequations}
\begin{eqnarray}
\label{rotscala}
C_u(\tau)_{R_e\rightarrow\infty} &\rightarrow& \frac{8\sqrt{2}\pi^2q\sqrt{3}}{\sqrt{2-q}}\int_{k_0}^{k_m} \sinh\left(\frac{\sqrt{2(2-q)}}{q\sqrt{3}}\tau\right)
\frac{dk}{k^2}\\
\label{rotscalb}
&\sim&\tau,\,\,\,{\rm when}\,\,\,q\rightarrow 2\,\,\,{\rm or}\,\,\,\tau\rightarrow 0,\\
\label{rotscalc}
{\rm but}\,\,\,&\sim&\exp\left(\frac{\sqrt{2(2-q)}}{q\sqrt{3}}\tau\right)\,\,\,{\rm for}\,\,\,{\rm large}\,\,\,\tau.
\end{eqnarray}
\label{rotscal}
\end{subequations}
Hence, the correlation effectively appears independent of $q$ at small and large $\tau$, at a large $R_e$ in 
particular. However, at an intermediate $\tau$, correlation diverges at $q\rightarrow 2$ --- when the 
background specific angular momentum is conserved. Then with the decrease of $q$, due to appearance of fluctuation due to rotational/Coriolis effects with nonzero epicyclic frequency (see, \cite{man,amn}
for a detailed discussion),
correlation decreases (see Fig. \ref{tempr1e4} also). However, the correlation is effectively controlled
by the dynamics of the flow, independent of viscosity.

For completeness we also calculate the correlation for nonrotating flows, given by
\begin{subequations}
\begin{eqnarray}
\label{norotscala}
\hskip0.2cm
C_u(\tau)_{\rm nonrot}&=&8\sqrt{2}\pi^2R_e\int_{k_0}^{k_m}\frac{\sinh\left(\frac{k^2\tau}{R_e}\right)}{k^4}\,\,dk\\
\label{norotscalb}
&\sim&\tau\,\,\,{\rm when}\,\,\,\frac{\tau}{R_e}<<1\\
\label{norotscalc}
&\sim&R_e\,\,\,{\rm when}\,\,\,R_e,\tau\rightarrow\infty\\
\label{norotscald}
&\sim&\frac{\tau^{3/2}}{R_e^{1/2}}\,{\rm erfi}\left(k\sqrt{\frac{\tau}{R_e}}\right)|_{k_0}^{k_m}\,\,\,{\rm when}
\,\,\tau\rightarrow\infty\,\,
{\rm or}\,\,\frac{\tau}{R_e}>>1.
\end{eqnarray}
\label{nonrotscal}
\end{subequations}
Hence, unless $\tau$ is very small at a finite $R_e$, correlation depends on $R_e$ --- quite an opposite trend
of the rotating cases.
Interestingly in absence of pressure, $C_u(\tau)_{\rm nonrot}\sim\tau^3$ as $\tau\rightarrow\infty$.
Therefore, the rotation and pressure both create significant impact in the behavior 
in the flow such that the scaling behavior changes.

Once $u_{\rm rms}$, $\frac{\partial u_{\rm rms}}{\partial {\vec x}}$ and $\zeta_{\rm rms}$ (when
${A_{\rm rms}}^2=|< {[A(\vec x,t+\tau)-A(\vec x,t)]}^2 >| \sim 
|< A(\vec x,t)\, A(\vec x,t+\tau) >|$, where $A\equiv u,\,\frac{\partial u}{\partial\vec x},\,\zeta$)
are known, the evolution of perturbation energy growth 
$\mathcal{E} = \frac{1}{8}[{u_{\rm rms}}^2 + \frac{1}{k^2}{(\frac{\partial u_{\rm rms}}{\partial {\vec x}})}^2 + {\zeta_{\rm rms}}^2]$
can be computed.
However, by simple calculation of scaling law, it can be shown that each
of the terms in $\mathcal{E}$ scales with $\tau$ identically and hence the scaling
behavior of $\mathcal{E}$ is same as that of $C_u(\tau)$ apart
from a constant factor. 

\begin{figure}
\vskip 0.3in
   \centering
\includegraphics[scale=0.8]{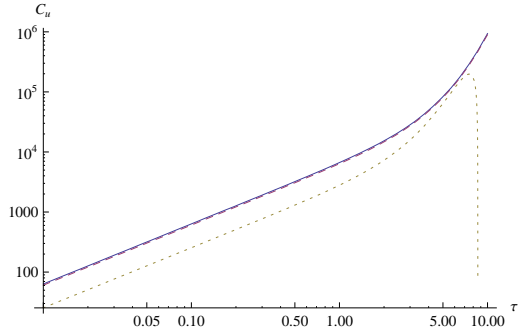}
\vskip0.5cm
\caption{Temporal correlation as a function of time for $q=1.5$, when $R_e=10000$ (solid line), $1000$ (dashed line) and $100$ (dotted line).
}
\label{tempq1p5}
\end{figure}

\begin{figure}
\includegraphics[scale=0.8]{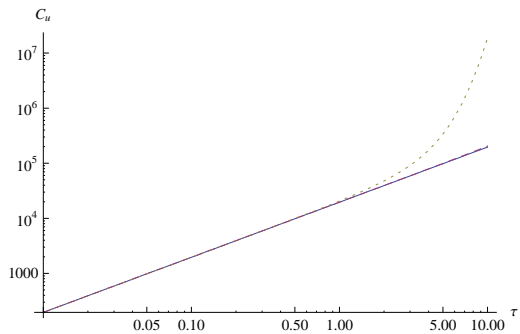}
\vskip0.5cm
\caption{Same as in Fig. \ref{tempq1p5}, except for non-rotating flow.
}
\label{tempnorot}
\end{figure}

\subsubsection{Numerical solutions}
Figure \ref{tempr1e4} shows that with the decrease of $q$, temporal correlation of velocity
perturbation decreases as the fluctuation due to the presence of Coriolis force
increases. At small values of $q$ (e.g. for $1,1.5$), however, the Coriolis fluctuation interacting with 
noise appears to be dominating in a way that 
the correlation reveals an oscillatory amplitude (this is more clear from Fig.
\ref{tempq1p5}). Hence in a few rotation time the correlation exceeds that of 
large $q$, when it increases steadily in absence of dominant Coriolis effect at large $q$. However, in 
either of the regimes, temporal correlation of perturbation appears large enough in a few rotation time, which reveals nonlinearity in the 
system, presumably leading to instability.
The correlation appears to be increasing steadily for the nonrotating case, in absence of Coriolis
fluctuation, similar to that for $q\rightarrow 2$.
Thus the interesting outcome here is the following. The Coriolis fluctuations along with a steady 
effect of noise lead to a larger $C_u(\tau)$ for a smaller $q$ than
that of a larger $q$ in a few rotation time, as is also indicated by equation (\ref{rotscal}). 
Note that an important physical quantity in this context is the 
rotation time of the flow. In a few rotation time, noise and the Coriolis
fluctuation together bring in a large $C_u(\tau)$ and hence perturbation energy (prone to a 
smaller $q$, such as $1.5$), which presumably could lead to a transition
to turbulence. Therefore,
the presence of noise makes the Coriolis effect active to generate instability, 
which were otherwise acting as the agents hindering 
the perturbation effects (see, e.g. \cite{man,amn}), leading to a faster rate
of perturbation energy growth into the system.

Figure \ref{tempq1p5} shows that for $q<2$, the correlation tends to become independent 
of $R_e$. This is because with decreasing $q$, the epicyclic frequency ($\kappa=\Omega\sqrt{2(2-q)}$) becomes larger
bringing lots of fluctuation in the flow, which makes the flow
limited by the dynamics itself but not by the viscosity. In absence of noise, perturbation 
energy growth (and hence
$C_u(\tau)$) is very small, as shown previously \cite{man}, for $q<2$, particularly in three 
dimensions. 
Note, however, that at a small/moderate $R_e$, epicyclic
fluctuations work with noise in such a way that $C_u(\tau)$ at small $\tau$ appears smaller than that of larger $R_e$ 
(see $R_e=100$ case in 
Fig. \ref{tempq1p5}) due to the harmonic nature of fluctuations. This effectively argues that 
the correlation saturates only with the increase of $R_e$ for $q<2$.

Figure \ref{tempnorot} shows
that $C_u(\tau)$ for a nonrotating flow at $R_e=100$ appears larger at large $\tau$ compared to that at
larger values of $R_e$. This is due to the effect of noise, which would have been suppressed at larger $R_e$. 
This is also clearly understood from 
the analytical scaling laws given by equation (\ref{nonrotscal}). 
Note that, a very large 
$R_e$ along with a large $\tau$ diminishes
fluctuations in the flow leading to a linear perturbation growth with $R_e$, as shown in equation (\ref{norotscalc}).

Comparing rotating and non-rotating cases, it can be further inferred that
the noise suppresses the viscous effects in the rotating 
flow making it independent of $R_e$ in the regime of smaller $\tau$, as also shown by equation 
(\ref{rotscalb}). 
However, the optimum perturbation of a flow might reveal the
correlation varying with $R_e$, and the above inferences are
valid for a fixed perturbation, considered for all the present cases.

Note that in absence of noise, perturbation energy growth (and hence
$C_u(\tau)$) in all the above rotating cases, essentially for $q<2$, is very small, as shown previously \cite{man}, 
particularly in three dimensions. However, the effects of noise bring
a steady growth in the perturbation top of the Coriolis fluctuations. 
A remarkable feature in the scaling nature of all these correlations, atleast for asymptotic $\tau$ (see, e.g., equations (\ref{rotscalb}) and (\ref{rotscalc})), 
is their independence of $q$ (background angular velocity profile) --- 
a trait identified 
in statistical physics literature as {\it universality}. All these flows, 
with identical overall perturbation energy growth 
exponents \cite{barabasi_stanley} in asymptotic limits of $\tau$ (and also $R_e$), but generally with varying 
energy dissipation amplitudes, indicate a specific universality class for them.

\subsection{Spatial correlation}

Here also we assume
$<\tilde{{\eta}_1}_{\vec{k},\omega}\,\tilde{{\eta}_2}_{\vec{k},\omega}>=0$ like temporal correlation and 
obtain spatial correlations of
velocity, vorticity and gradient of velocity, given below as
\begin{eqnarray}
\nonumber
&&<u(\vec{x},t)\,u(\vec{x}+\vec{r},t)>=S_u(r)=\int d^3k\,d\omega\,e^{i\vec k.\vec r}
<\tilde{u}_{\vec{k},\omega}\,
\tilde{u}_{-\vec{k},-\omega}>,\\
\nonumber
&&<\zeta(\vec{x},t)\,\zeta(\vec{x}+\vec{r},t)>=S_\zeta(r)=\int d^3k\,d\omega\,e^{i\vec k.\vec r}
<\tilde{\zeta}_{\vec{k},\omega}\,
\tilde{\zeta}_{-\vec{k},-\omega}>,\\
&&{\hskip-1.3cm <\frac{\partial u(\vec{x},t)}{\partial x}\,\frac{\partial u(\vec{x}+\vec r,t)}
{\partial x}>
=S_{du}(r)
=\int d^3k\,d\omega\,e^{i\vec k.\vec r}k_x^2<\tilde{u}_{\vec{k},\omega}\,
\tilde{u}_{-\vec{k},-\omega}>},
\label{velzetdercor2}
\end{eqnarray}
where $\vec{r}$ be the position where the correlations to be measured.

Now using equations (\ref{mat1}), (\ref{mat2}), (\ref{am}) and (\ref{velzetdercor2}), the 
velocity correlation perturbation $S_u(r)$ 
is explicitly given by

\begin{eqnarray}
&&S_u(r) 
\nonumber
= 2\pi\int_{k_0}^{k_m}~dk~k^2 ~\int_0^\pi~d\theta~e^{ikr\cos\theta}~
\int d\omega~[2D_1 q^4 R_e^2 \\ 
\nonumber
&&(k^4+(R_e \omega + LR_e k_y)^2) + 8D_2 q^2 R_e^4 k_z^2]/
[4 (2-q)^2 R_e^4 k_z^4 +\\
\nonumber
&&4 R_e^2 q^2 (2-q) k^2 k_z^2 (k^4-(R_e \omega + LR_e k_y)^2) 
+ k^4 q^4 (k^4+\\
\nonumber
&&{(R_e \omega + LR_e k_y)}^2)^2] \\ \nonumber
&&= 2\pi^2~\int_{k_0}^{k_m}~dk~k^2~J_0(kr)~\int d\omega~[2D_1 q^4 R_e^2 (k^4 +(R_e 
\omega + LR_e k_y)^2)\\
\nonumber 
&&+ 8D_2 q^2 R_e^4 k_z^2]/[4 (2-q)^2 R_e^4 k_z^4 +4 R_e^2 q^2 (2-q) k^2 k_z^2 (k^4-(R_e \\
&&\omega + LR_e k_y)^2) + k^4 q^4 (k^4+(R_e \omega + LR k_y)^2)^2],
\label{spacecorr}
\end{eqnarray}

\noindent
where $J_0(kr)$ is the zeroth-order Bessel function. Similarly, one can
obtain $S_\zeta(r)$ and $S_{du}(r)$ explicitly, when the poles in equation 
(\ref{spacecorr}) are exactly the same as that given by equation (\ref{poles}). 
Here also we stick to the simplifying assumption $k_x=k_y=k_z=k/\sqrt{3}$. 

Now following the same procedure as applied for the correlation of temporal velocity perturbation, 
we obtain 
the correlation of spatial velocity perturbation $S_u(r)$ given by
\begin{eqnarray}
&&S_u(r) = \int_{k_0}^{k_m}~dk~k^2~J_0(kr)~{\rm RES}_{S_u}(q,R_e,r), 
\label{spacecorr2}
\end{eqnarray}
where ${\rm RES}_{S_u}(q,R_e,r)$ is the magnitude of the part arising from frequency integration.

\subsubsection{Asymptotic scaling laws}

At large $r$ and 
$R_e$ (which is a natural/astrophysical scenario), the flows with all $q$ reveal 
identical scaling of correlation with $r$ given by
\begin{eqnarray}
{S_u}_{(R_e\,{\rm and}\,r\rightarrow\infty)}\rightarrow\frac{\pi^{3/2} q}{r^{1/2}}\,(2-q)^{-3/2}\,\int_{k_0}^{k_m}dk\,\exp(-kr)
\, \frac{2\sqrt{3}(k^2-1)+q\sqrt{3}}{k^{5/2}}
\label{spatinf}
\end{eqnarray}
which is independent of $R_e$. Note that at a finite $r$, $\exp(-kr)/\sqrt{r}\rightarrow J_0(kr)$. 
Hence, the flow is limited by dynamics.
This reveals identical roughness exponents \cite{barabasi_stanley} for all flows, indicating a
single universality class, as was already predicted from the temporal correlations.
Equation (\ref{spatinf}) also reveals a steadily damped instability. It further 
recovers the fact that the flow exhibits unbounded correlation as $q\rightarrow 2$,
and hence unbounded growth of perturbation. At small $r$ 
\begin{eqnarray}
\nonumber
{S_u}_{(R_e\rightarrow\infty,\,\,r\rightarrow 0)}\rightarrow -q\,(2-q)^{-3/2}\int_{k_0}^{k_m}dk\,
\frac{(2-q-2k^2)\log\left(\frac{kr}{2}\right)}{k^2},\\
\label{spat0}
\end{eqnarray}
which is again determined by rotational parameter $q$ only, when $\log(kr/2)$ becomes negative
at a small $r$. Hence, except $q=2$ (or $q$ very very close to $2$) and/or
$r=0$ (or $r$ very very close to $0$), the correlation is finite.

\begin{figure}
\vskip 0.3in
   \centering
\includegraphics[scale=0.8]{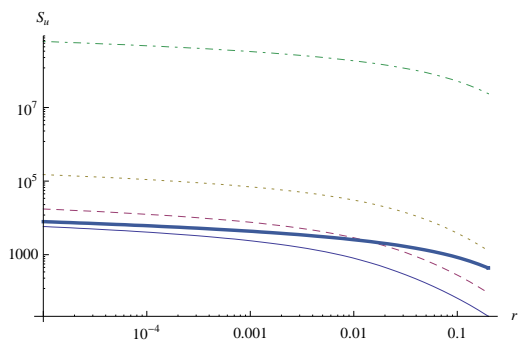}
\vskip0.5cm
\caption{Spatial correlation as a function of radial coordinate for $R_e=10000$, when $q=1$ (solid line), $1.5$ 
(dashed line), $1.9$ (dotted line), $1.9999$ (dotdashed line) and
flow is non-rotating (thick line).
}
\label{spatr1e4}
\end{figure}

\begin{figure}
\vskip 0.3in
   \centering
\includegraphics[scale=0.8]{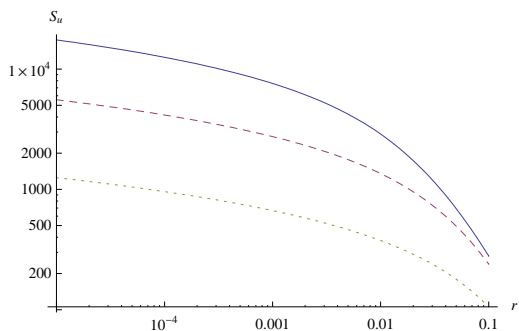}
\vskip0.5cm
\caption{Spatial correlation as a function of radial coordinate for $q=1.5$, when $R_e=10000$ (solid line), 
$1000$ (dashed line) and $100$ (dotted line).
}
\label{spatq1p5}
\end{figure}

\begin{figure}
\vskip 0.3in
   \centering
\includegraphics[scale=0.8]{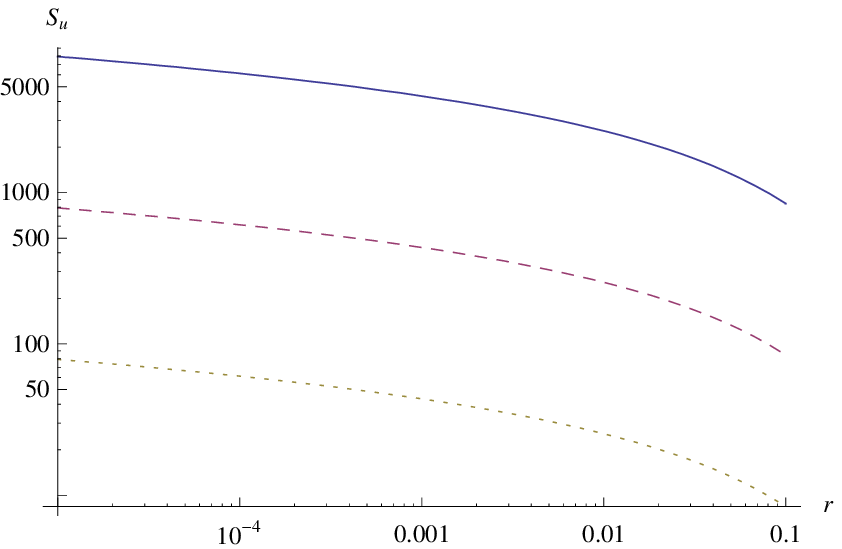}
\vskip 0.5cm
\caption{Same as in Fig. \ref{spatq1p5}, except for non-rotating flow.
}
\label{spatnorot}
\end{figure}

For completeness, we also describe scaling laws for nonrotating flows.
At large $r$ and $R_e$
\begin{eqnarray}
{S_u}_{(R_e\,{\rm and}\,r\rightarrow \infty)}^{\rm nonrot}\rightarrow 
\frac{R_e\,\pi^{3/2}\sqrt{2}}{r^{1/2}}\int_{k_0}^{k_m}dk\,
\frac{\exp(-kr)}{k^{9/2}},
\label{spatinfrot0}
\end{eqnarray}
which scales with $R_e$, unlike the rotating flows. Hence the flow is not limited by dynamics
at any $r$. However, at $r\rightarrow 0$ the scaling behavior 
changes and is given by
\begin{eqnarray}
{S_u}_{(R_e\rightarrow\infty,\,\,r\rightarrow 0)}^{\rm nonrot}\rightarrow -2R_e\pi\int_{k_m}^{k_0}dk\,
\frac{\log\left(\frac{kr}{2}\right)}{k^4},
\label{spatrot0}
\end{eqnarray}
which still scales linearly with $R_e$.

\subsubsection{Numerical solutions}

Figures \ref{spatr1e4}-\ref{spatq1p5} show that the spatial correlations of perturbation
decrease with the decrease of $q$ from $2$, when the epicyclic fluctuations arise into the flow.
Moreover, the correlations for all $R_e$ tend to merge at large $r$, at a given $q$ particularly, 
becoming independent of $R_e$,
as is also shown by the asymptotic result given by equation (\ref{spatinf}).  
However, at small $r$ and finite $R_e$, they depend on viscosity.
This is due to the effect of noise, as is the case for temporal correlations.

On the other hand, for non-rotating flows, the correlations are not
limited by the dynamics of the flow, as shown by Fig. \ref{spatnorot}, rather
are controlled by viscosity. This is also understood from equations (\ref{spatinfrot0}) and (\ref{spatrot0}). 
Hence, the rotational effects constraint the correlation noticeably.

It is generally seen that the correlations decrease with increasing $r$ and decreasing $q$.
Note their exponential decaying nature in equations (\ref{spatinf}) and (\ref{spatinfrot0}) 
compared to the logarithmic variance in equations (\ref{spat0}) and (\ref{spatrot0}). 
However, their value appears significant enough
to reveal a steadily damped instability in the flow. 
The possibilities of steadily damped oscillatory instabilities in 
constrained Orr-Sommerfeld 
flows were already predicted in two dimensions \cite{akhavan91} as well in the recirculating cavities 
between the jets produced behind a high solidity 
grid consisting of a perforated plate \cite{villermaux91}. 

\begin{figure}
\vskip 0.3in
   \centering
\includegraphics[scale=0.8]{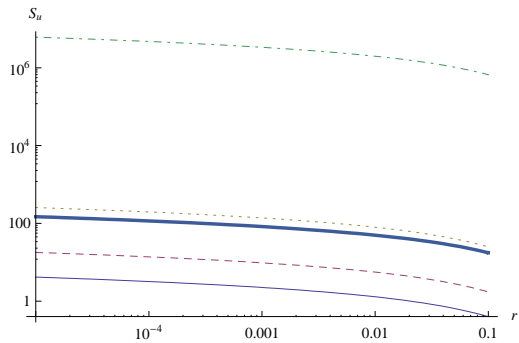}
\vskip0.5cm
\caption{Same as in Fig. \ref{spatr1e4}, except for $D_1=D_2\sim 1/k^3$.
}
\label{spatr1e4vd}
\end{figure}

\begin{figure}
\vskip 0.3in
   \centering
\includegraphics[scale=0.8]{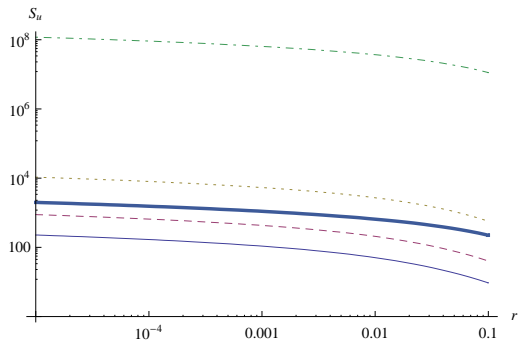}
\vskip0.5cm
\caption{Same as in Fig. \ref{spatr1e4}, except for $D_1=D_2\sim 1/k$.
}
\label{spatr1e4vd2}
\end{figure}

All the above correlations are computed for white noise. Figures \ref{spatr1e4vd} and 
\ref{spatr1e4vd2} show the variations of spatial correlations of velocity perturbation for color noise
respectively without and with vertex corrections in $D_i(k)$. Interestingly, while the
magnitude of correlations decreases with respect to that for white noise (particularly
for $D_1=D_2\sim 1/k^3$), with the increase of $q$ they appear large enough to govern
instability. As an example, for noise with vertex correction, the Keplerian flow still reveals
large enough correlation even at small radial coordinates. \\ \\

All the above large values of perturbation energy growth are indicative of pure hydrodynamic
instability and plausible turbulent transport, which is one of the outcomes under this 
theory. 


\subsection{Spectrum}

\begin{figure}
\vskip 0.3in
   \centering
\includegraphics[scale=0.9]{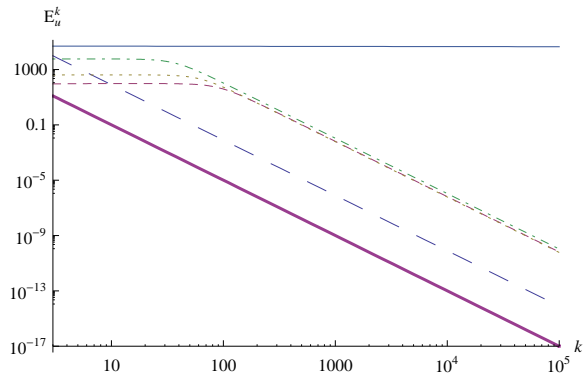}
\vskip0.5cm
\caption{Energy spectra for the velocity perturbation, when $q=1$ (dashed line), $1.5$ (dotted line) and 
$1.9$ (dotdashed line) and flow is nonrotating (longdashed). The solid and thick-solid lines respectively
indicate the nature of slop of the curves at small and large $k$ for rotating flows. $D_1=D_2=1$, $R_e=10000$.
}
\label{specr1e4vd}
\end{figure}

\begin{figure}
\vskip 0.3in
   \centering
\includegraphics[scale=0.9]{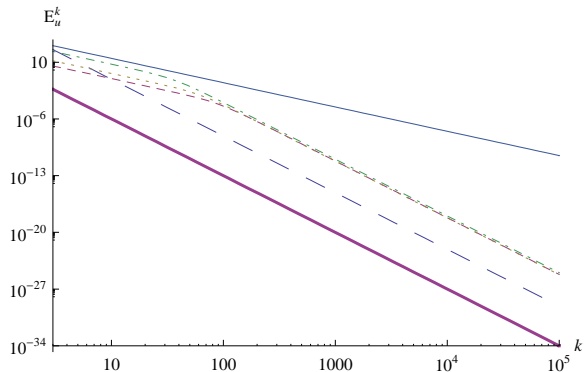}
\vskip0.5cm
\caption{Same as in Fig. \ref{specr1e4vd}, except for $D_1=D_2\sim 1/k^3$.
}
\label{specr1e4vd2}
\end{figure}

In order to compute the spectrum of flow, let us first integrate $\omega$-part of 
equation (\ref{spacecorr}) and obtain
\begin{eqnarray}
\int E_u(k)~dk=\int d^3k~{\rm RES}_{S_u}(q,R_e,k,D_1,D_2).
\end{eqnarray}
Similarly, one can obtain $\int E_\zeta(k)dk$ and $\int E_{du}(k)dk$ from $S_{\zeta}$ 
and $S_{du}$ respectively.
They would scale in a same way as of $\int E_u(k)dk$.
Hence, the energy spectrum can generally be given by
\begin{eqnarray}
E_u(k)=C~k^2~{\rm RES}_{S_u}(q,R_e,k,D_1,D_2),
\label{spec}
\end{eqnarray}
when $C$ is a constant. Figures \ref{specr1e4vd} and \ref{specr1e4vd2} show that the energy spectra 
for all the rotating flows with $q<2$ have same spectral indices at large and small $k$. Moreover,
the energy appears identical for all $q$ at large $k$, which is due to the fact that a small length scale
appears identical for all rotational profiles. At large scale, the spectral index is same
as that of noise correlation, which implies that the flow is controlled by noise in that regime. 
However, interestingly, there
is no change of slop in the nonrotating case, and the corresponding power-law index appears 
same as that of the large $k$ rotating ones. This may be understood as the Coriolis
force in the rotating flows distorts the system which results in affecting the large scale
(and hence small $k$) behavior of the flow. Hence, the behavior of non-rotating flow appears
same throughout without being affected by the rotational effects. Note also that the color
noise increases the power-law indices throughout compared to that for white noise.

\section{Summary and conclusion}

We have attempted to address the origin of instability and then turbulence in rotating shear flows (more precisely
a small section of it, which is a plane shear flow with the Coriolis force). Our particular emphasis is
the flows having angular velocity decreases but specific angular momentum increases with the radial
coordinate, which are Rayleigh stable. The flows with such a kind of velocity profiles are often seen in astrophysics. 
As the molecular viscosity in astrophysical accretion disks
is negligible, any transport of matter therein would arise through turbulence. However, the
Rayleigh stable flows, particularly in absence of magnetic coupling, could not reveal any
unstable mode which could serve as a seed of turbulence. Therefore, essentially we have addressed the
plausible origin of {\it pure hydrodynamic turbulence} in rotating shear flows of the kind mentioned above. 
This is particularly meaningful for charge natural flows like accretion disks around quiescent cataclysmic variables, 
in protoplanetary and star-forming disks, 
and the outer region of disks in active galactic nuclei etc. where flows are cold and of
low ionization and effectively neutral in charge.
Note that whether a flow is magnetically arrested or not, the magnetic instability is there or not,
hydrodynamic effects always exist. Hence, the relative strengths of hydrodynamics and hydromagnetics 
in the time scale of interest determines the actual source of instability.

We have shown, based on the theory of statistical physics, that stochastically forced linearized rotating shear flows 
in a narrow gap limit reveal a large correlation energy growth of
perturbation in presence of noise.
While the correlations are directly proportional to the Reynolds number for nonrotating flows,
they are independent or weakly dependent of Reynolds number for rotating flows. 
Hence the later flows are mostly controlled by the dynamics
and noise. While correlations of perturbation decrease as the flow deviates
from the type of constant specific angular momentum to of the Keplerian and then to of constant circular velocity,
still they appear large enough to trigger nonlinear effects and instability. 

While earlier the origin of hydrodynamic instability in rotating shear flows of said kind 
was addressed based on ``bypass method" in two dimensions \cite{man,amn},
it was shown that \cite{bm06} extra physics, e.g. elliptical vortex effects, has to be 
invoked for such instabilities to occur in three dimensions. However, later theories showing
instabilities are effectively based on nonlinear theory. 
Therefore, the present work addresses the 
{\it large three-dimensional hydrodynamic energy growth of linear perturbation} 
for the first time, in such rotating flows, to best of our knowledge, which
presumably leads to instability and subsequent turbulence in such flows.
Only requirement here
is to have the presence of stochastic noise in the system, which is quite obvious in natural flows 
like astrophysical accretion disks. Indeed, to best of our knowledge, no one so far
has addressed the effects of stochastic noise in such rotating flows.

Interestingly, the flows with velocity profile $\Omega\sim r^{-q}$, with $q<2$, exhibiting identical overall 
growth exponents (but may not be identical energy dissipation amplitudes), indicate a specific 
universality class.
At large $r$, they also reveal identical roughness exponents.
Thus the properties of temporal and spatial correlations together indicate 
the Rayleigh stable rotating shear flows to follow a single universality class. 

Let us end by clarifying suitability of astrophysical application of our results
which is based on incompressible assumption.
If the wavelength of the velocity perturbations is much shorter
than the sound horizon for the time of interest (e.g. infall time of matter), then
the density perturbations (i.e. sound waves) reach
equilibrium early on, exhibiting effectively
uniform density during the timescale of interest. 
For a geometrically thin astrophysical accretion disk around a
gravitating mass, the vertical half-thickness
of the disk ($h$) is comparable to the sound horizon
corresponding to one disk rotation time
\cite{prin} such that $h \sim c_s / \Omega$. 
Therefore, as described in previous work \cite{chage,yeko,amn,jg05,umregmen,rajesh}, for 
processes taking longer than one rotation time,
wavelengths shorter than the disk thickness can be
approximately treated as incompressible. A detailed astrophysical application of the present
work has been reported elsewhere \cite{mukhch}.

\section*{Acknowledgments}
The authors would like to thank Rahul Pandit for discussion. Thanks are also due to
the referees for carefully, thorough reading the paper and various suggestions 
which have helped to improve the presentation of the work.
This work is partly supported by a project, Grant No. ISRO/RES/2/367/10-11, funded
by ISRO, India.

\appendix
\section*{Appendix: Magnetic version of the set of Orr-Sommerfeld and Squire equations}

Let us now establish the linearized Navier-Stokes equation in presence of background plane shear $(0,-x,0)$ without
neglecting magnetic coupling with a background magnetic field $(0,B_1,1)$,
when $B_1$ being a constant, in presence of angular velocity 
$\Omega\sim r^{-q}$ in a small section of the incompressible flow. Hence, the underlying equations
are nothing but the linearized set of hydromagnetic equations including the equations of induction in 
local Cartesian coordinates (see, e.g., \cite{man,bmraha} for detailed description of the
choice of coordinate in a small section) given by
\begin{eqnarray}
\left(\frac{\partial }{\partial t}-x\frac{\partial }{\partial y}\right)u-\frac{2v}{q}+\frac{\partial p_{\rm tot}}
{\partial x}-\frac{1}{4\pi}\left(B_1\frac{\partial B_x}{\partial y}+\frac{\partial B_x}{\partial z}\right)
=\frac{1}{R_e}\nabla^2u,
\label{u}
\end{eqnarray}
\begin{eqnarray}
\left(\frac{\partial }{\partial t}-x\frac{\partial }{\partial y}\right)v+\left(\frac{2}{q}-1\right)u+\frac{\partial p_{\rm tot}}
{\partial y}-\frac{1}{4\pi}\left(B_1\frac{\partial B_y}{\partial y}+\frac{\partial B_y}{\partial z}\right)
=\frac{1}{R_e}\nabla^2v,
\label{v}
\end{eqnarray}
\begin{eqnarray}
\left(\frac{\partial }{\partial t}-x\frac{\partial }{\partial y}\right)w+\frac{\partial p_{\rm tot}}
{\partial z}-\frac{1}{4\pi}\left(B_1\frac{\partial B_z}{\partial y}+\frac{\partial B_z}{\partial z}\right)
=\frac{1}{R_e}\nabla^2w,
\label{w}
\end{eqnarray}
\begin{eqnarray}
\frac{\partial B_x}{\partial t}=\frac{\partial u }{\partial z}+ B_1\frac{\partial u}{\partial y}+x\frac{\partial B_x}
{\partial y}+\frac{1}{R_m}\nabla^2B_x,
\label{bx}
\end{eqnarray}
\begin{eqnarray}
\frac{\partial B_y}{\partial t}=\frac{\partial v }{\partial z}+ B_1\frac{\partial v}{\partial y}-x\frac{\partial B_x}
{\partial x}-x\frac{\partial B_z}{\partial z}-B_x+\frac{1}{R_m}\nabla^2B_y,
\label{by}
\end{eqnarray}
\begin{eqnarray}
\frac{\partial B_z}{\partial t}=\frac{\partial w }{\partial z}+ B_1\frac{\partial w}{\partial y}+x\frac{\partial B_z}
{\partial y}+\frac{1}{R_m}\nabla^2B_z,
\label{bz}
\end{eqnarray}
when the vectors for velocity and magnetic field perturbations are $(u,v,w)$ and $(B_x,B_y,B_z)$ respectively,
$R_e$ and $R_m$ are the hydrodynamic and magnetic Reynolds numbers respectively, $p_{\rm tot}$ is the total
pressure perturbation (including that due to the magnetic field). Above equations are
supplemented by the conditions for incompressibility and no magnetic charge, given by
\begin{eqnarray}
\frac{\partial u}{\partial x}+\frac{\partial v }{\partial y}+ \frac{\partial w}{\partial z}=0,
\label{eoc}
\end{eqnarray}
\begin{eqnarray}
\frac{\partial B_x}{\partial x}+\frac{\partial B_y }{\partial y}+ \frac{\partial B_z}{\partial z}=0.
\label{monop}
\end{eqnarray}
Now by taking partial derivatives with respect to $x,y,z$ respectively to both the sides of equations (\ref{u}),
(\ref{v}), (\ref{w}) and adding them up we obtain 
\begin{eqnarray}
\nabla^2p_{\rm tot}=\frac{2}{q}\left(\frac{\partial v}{\partial x}-\frac{\partial u}{\partial y}\right)+
2\frac{\partial u}{\partial y}.
\label{delp}
\end{eqnarray}
Now taking $\nabla^2$ to equation (\ref{u}), using equation (\ref{delp}) and defining $x$-component of vorticity
$\zeta={\partial w}/{\partial y}-{\partial v}/{\partial z}$, we obtain
\begin{equation}
\left(\frac{\partial}{\partial t}-x\frac{\partial}{\partial y}\right)\nabla^2 u
+\frac{2}{q}\frac{\partial \zeta}{\partial z}-\frac{1}{4\pi}\left(B_1\frac{\partial}{\partial y}
+\frac{\partial}{\partial z}\right)\nabla^2B_x
=\frac{1}{R_e}\nabla^4 u.
\label{orr}
\end{equation}
Also by taking partial derivatives with respect to $x$ and $y$ respectively to both the sides of equations (\ref{v}),
(\ref{u}) respectively, subtracting one from other and defining 
$\zeta_B={\partial B_z}/{\partial y}-{\partial B_y}/{\partial z}$, we obtain
\begin{equation}
\left(\frac{\partial}{\partial t}-x\frac{\partial}{\partial y}\right)\zeta+
\left(1-\frac{2}{q}\right)\frac{\partial u}{\partial z}-\frac{1}{4\pi}\left(B_1\frac{\partial}{\partial y}
+\frac{\partial}{\partial z}\right)\zeta_B
=\frac{1}{R_e}\nabla^2 \zeta.
\label{orr2}
\end{equation}
Furthermore, by taking partial derivatives with respect to $y$ and $z$ respectively to both the sides of equations 
(\ref{bz}) and (\ref{by}) and subtracting one from other, we obtain
\begin{eqnarray}
\left(\frac{\partial }{\partial t}-x\frac{\partial}{\partial y}\right)\zeta_B=
\frac{\partial \zeta}{\partial z}+B_1\frac{\partial \zeta}{\partial y}+
\frac{\partial B_x}{\partial z}+\frac{1}{R_m}\nabla^2\zeta_B.
\label{bzet}
\end{eqnarray}
The equations (\ref{orr}), (\ref{orr2}), (\ref{bx}) and (\ref{bzet}) describe the set of magnetized
version of the Orr-Sommerfeld and Squire equations in presence of the Coriolis force, for the background 
magnetic field described above and linear shear. To best of our knowledge, this set of 
equations has not been 
reported in the existing literature so far.

{}

\end{document}